\documentclass[aps,rpl,preprint,groupedaddress]{revtex4}
\begin{document}

\title{Effects of Collisional Decoherence on Multipartite Entanglement - How would entanglement not be relatively common?}

\author{Guang-Hao Low}
\affiliation{Hwa Chong Institution, 661 Bukit Timah Road, Singapore 269734, Singapore}

\author{Zhiming Shi}
\affiliation{Hwa Chong Institution, 661 Bukit Timah Road, Singapore 269734, Singapore}

\author{Ye Yeo}
\affiliation{Department of Physics, National University of Singapore, 10 Kent Ridge Crescent, Singapore 119260, Singapore}

\begin{abstract}
We consider the collision model of Ziman {\em et al.} and study the robustness of $N$-qubit Greenberger-Horne-Zeilinger (GHZ), W, and linear cluster states.  Our results show that $N$-qubit entanglement of GHZ states would be extremely fragile under collisional decoherence, and that of W states could be more robust than of linear cluster states.  We indicate that the collision model of Ziman {\em et al.} could provide a physical mechanism to some known results in this area of investigations.  More importantly, we show that it could give a clue as to how $N$-partite distillable entanglement would be relatively rare in our macroscopic classical world.
\end{abstract}

\maketitle

Entanglement lies at the heart of quantum information theory \cite{Nielsen}.  It has recently been studied extensively in the context of quantum computation, which would require the creation and maintenance of highly complex entangled states of $N$ particles, $A_1,\ \cdots,\ A_N$, where $N$ can be arbitrarily large.  This is a difficult task because unavoidable random interactions of the composite system $A_1\cdots A_N$ with its environment creates entanglement between them, which at the same time degrades the entanglement within the system itself.  Active intervention in the form of distillation protocols and error correcting codes is capable of maintaining the $N$-partite entanglement in the presence of {\em decoherence}.  However, the efficiency of such methods expectedly depends on the {\em a priori} robustness of the entangled state in question, i.e., the ability of an entangled state to remain entangled under ({\em local}) decoherence.  Local here means that individual particle $A_i$ of the system interacts independently with the environment.  Several natural questions of practical importance are thus: What is the robustness of multipartite entanglement under certain decoherence model?  How the robustness of the entanglement changes with the size $N$ of the system?  How the robustness of one multipartite entangled state compares with other multipartite entangled states?

In order to gain some insight into the above questions, Simon and Kempe \cite{Simon} were the first to analyze the entanglement properties of $N$-qubit Greenberger-Horne-Zeilinger (GHZ) states \cite{Greenberger} under the action of ({\em homogeneous}) local depolarizing channels.  They studied how much local depolarization is possible such that the states are still entangled.  They made comparisons with the W states \cite{Zeilinger}, {\em albeit} for $N = 3, 4$.  Later, Bandyopadhyay and Lidar \cite{Bandyo} considered the robustness of $N$-qubit GHZ states under local generalized depolarizing channels.  Carvalho, Mintert and Buchleitner \cite{Carvalho} extended the above studies of the $N$-qubit GHZ and W states to situations involving the presence of local amplitude damping and dephasing channels.  In \cite{DurII, Hein}, Dur, Briegel and Hein studied the entanglement properties of the rather general class of graph states, which includes GHZ and cluster states \cite{Briegel}, under local amplitude damping and generalized deoplarization.

The structure of $N$-qubit entanglement is considerably more complex than bipartite entanglement.  In particular, there is no such thing (yet) as a general practical criteria for multipartite entanglement (compare \cite{Peres, Wootters}).  So, to analyze the entanglement of the above multiqubit states, Simon and Kempe \cite{Simon} considered the entanglement for bipartite cuts, where the $N$ qubits are partitioned into two (arbitrarily chosen) subsystems, with some $n$ qubits constitute one subsystem ${\cal P}_1$ and all the other $N - n$ qubits the other subsystem ${\cal P}_2$.  For each such cut they employed the partial transposition criterion \cite{Peres}.  As long as the partial transpose of a state has negative eigenvalues, the state is definitely entangled.  It turns out that a necessary condition for $N$-qubit distillability is that the partial transpositions with respect to any group of qubits are nonpositive (see \cite{DurIII, DurIV}).  Consider $N$ distinct parties each holding a qubit belonging to the $N$-qubit state $\rho_{A_1\cdots A_N}$.  It is $N$-party distillable entangled if there exists a local protocol (i.e., the parties act individually on their qubits and are allowed to communicate classically) such that one can obtain from a sufficiently large number of copies of $\rho_{A_1\cdots A_N}$ some true $N$-qubit entangled pure state.  This distillable entanglement criterion is being used in \cite{DurII, Hein}.  It is of obvious practical relevance and will be the criterion employed in this paper.  We remark that Carvalho, {\em et al.}\cite{Carvalho} too used a different measure of multipartite entanglement - a specific generalization of concurrence \cite{Rungta} of pure bipartite states, that is sensitive to real multipartite correlations.

Carvalho, {\em et al.} \cite{Carvalho}  showed that only for depolarizing channels do the GHZ or W states turn separable after a finite time.  And, this occurs with the concurrence of the decaying GHZ state always larger than that of the decaying W state.  In contrast, the amplitude damping as well as the dephasing channels induce separability only in the limit of infinite time.  And, the concurrence of the decaying W state gets larger than that of the decaying GHZ state after a short time.  In terms of entanglement decay rates, they found that the GHZ states become separable at rates which increase linearly with $N$ (for $N \geq 4$).  For the W states, only the depolarizing channels give rise to an almost linear increase of entanglement decay rates with $N$, slightly faster than for the GHZ states.  The decay of the concurrence is independent of $N$ for amplitude damping and dephasing channels.  They concluded that the multipartite quantum correlations of W states outperform those of GHZ states in terms of their robustness.  Their results for the GHZ states agree with those in \cite{DurII, Hein}, which concluded via their analysis on linear cluster and more general graph states that true multipartite entanglement in macroscopic objects can be more stable and might be more common than previously thought \cite{CommentI}.

Recently, Ziman {\em et al.} \cite{ZimanI, ZimanII} showed that all {\em decoherence} maps of qubits, to be defined below, can be modelled as sequences of collisions between the system qubits under consideration and qubits from their environment \cite{CommentII}.  In this paper, we study the robustness of multiqubit GHZ, W and linear cluster states under collisional decoherence, which we briefly describe next.  We find that the $N$-qubit distillable entanglement associated with these states may not be robust, with the GHZ states the most fragile, and the linear cluster states less robust than W states.

Consider a system of $N$ qubits, $A_1, \cdots, A_N$, initially decoupled from an environment, i.e., the initial system-environment state is given by $\rho_{A_1\cdots A_N} \otimes \Xi_{env}$.  Following \cite{ZimanI}, we model the environment as a set of $\prod^N_{i = 1}N'_i$ qubits: $E_{ij}$, with $1 \leq i \leq N$ and $1 \leq j \leq N'_i$.  These environment qubits are initially in a factorized state:
$\Xi_{env} = \bigotimes^N_{i = 1}\bigotimes^{N'_i}_{j = 1}\xi_{E_{ij}}$, with $\xi_{E_{ij}} = \xi$ for all $E_{ij}$.  They do not interact between themselves, and each $E_{ij}$ undergoes a bipartite collision with each system qubit $A_i$ only once.  Under such conditions the evolution of each $A_i$ is described by the sequence of maps ${\cal E}^1_{A_i},\ \cdots,\ {\cal E}^{N'_i}_{A_i}$.  In particular, the state of $A_i$ after the $k_i$-th collision is given by $\rho^{(k_i)}_{A_i} \equiv {\cal E}^{k_i}_{A_i}[{\cal E}^{k_i - 1}_{A_i}[\cdots {\cal E}^1_{A_i}[\rho_{A_i}]\cdots]]$, where $\rho_{A_i} \equiv {\rm tr}_{A_1\cdots A_{i - 1}A_{i + 1}\cdots A_N}(\rho_{A_1\cdots A_N})$, ${\cal E}^j_{A_i}[\rho_{A_i}] = {\rm tr}_{E_{ij}}[U_{A_iE_{ij}}(\rho_{A_i} \otimes \xi_{E_{ij}})U^{\dagger}_{A_iE_{ij}}]$.  Here, the bipartite unitary operator that describes each collision,
\begin{equation}
U_{A_iE_{ij}} \equiv |0\rangle_{A_i}\langle 0| \otimes V^0_{E_{ij}} + |1\rangle_{A_i}\langle 1| \otimes V^1_{E_{ij}}.
\end{equation}
It belongs to the set of all controlled-$U$ transformations, where the system qubit $A_i$ plays the role of the control and the environment qubit $E_{ij}$ is a target.  With the unitary operators on $E_{ij}$ defined by
\begin{eqnarray}
V^0_{E_{ij}} & \equiv & |\psi\rangle_{E_{ij}}\langle 0| + |\psi^{\perp}\rangle_{E_{ij}}\langle 1|, \nonumber \\
V^1_{E_{ij}} & \equiv & |\phi^{\perp}\rangle_{E_{ij}}\langle 0| + |\phi\rangle_{E_{ij}}\langle 1|,
\end{eqnarray}
where $|\psi\rangle$ and $|\phi\rangle$ are arbitrary normalized kets with $\langle\psi^{\perp}|\psi\rangle = \langle\phi^{\perp}|\phi\rangle = 0$, it follows that the state of $A_i$ after the first collision is given by $\rho^{(1)}_{A_i}$ or
\begin{equation}
{\cal E}^1_{A_i}[\rho_{A_i}] = 
\rho^{00}_{A_i}P^0_{A_i} + \langle X_{E_{i1}}\rangle_{\xi}\rho^{01}_{A_i}S^+_{A_i} 
+ \langle X^{\dagger}_{E_{i1}}\rangle_{\xi}\rho^{10}_{A_i}S^-_{A_i} + \rho^{11}_{A_i}P^1_{A_i},
\end{equation}
where
$$
P^0 \equiv |0\rangle\langle 0| = \left(\begin{array}{cc} 1 & 0 \\ 0 & 0 \end{array}\right),
P^1 \equiv |1\rangle\langle 1| = \left(\begin{array}{cc} 0 & 0 \\ 0 & 1 \end{array}\right),
$$
\begin{equation}
S^+ \equiv |0\rangle\langle 1| = \left(\begin{array}{cc} 0 & 1 \\ 0 & 0 \end{array}\right),
S^- \equiv |1\rangle\langle 0| = \left(\begin{array}{cc} 0 & 0 \\ 1 & 0 \end{array}\right),
\end{equation}
such that $\rho_{A_i} = \rho^{00}_{A_i}P^0_{A_i} + \rho^{01}_{A_i}S^+_{A_i} + \rho^{10}_{A_i}S^-_{A_i} + \rho^{11}_{A_i}P^1_{A_i}$.  Here, the unitary operator $X_{E_{ij}} \equiv V^{1\dagger}_{E_{ij}}V^0_{E_{ij}}$, $\langle X^{\dagger}_{E_{ij}}\rangle_{\xi} = \langle X_{E_{ij}}\rangle^*_{\xi}$, and
\begin{equation}
\langle X_{E_{ij}}\rangle_{\xi} \equiv {\rm tr}[\xi X_{E_{ij}}] \equiv \lambda_{ij}\exp(I\phi_{ij}),
\end{equation}
with $0 \leq \lambda_{ij} \leq 1$ and $0 \leq \phi_{ij} \leq 2\pi$.  Whenever the collision is nontrivial, i.e., $V^1_{E_{ij}} \not= e^{-i\phi_{ij}}V^0_{E_{ij}}$, and the environment qubit state $\xi$ is not an eigenstate of $X_{E_{ij}}$, we have $0 \leq \lambda_{ij} < 1$ and Eq.(3) describes a quantum decoherence channel with decoherence basis $\{|0\rangle,\ |1\rangle\}$ \cite{ZimanI, ZimanII}.  From here on, we assume these conditions are fulfilled.  Specifically, after the $k_i$-th collision, we have $P^{0, 1}_{A_i} \longrightarrow P^{0, 1}_{A_i}$,
\begin{equation}
S^{\pm}_{A_i} \longrightarrow \gamma_i\exp(\pm I\Phi_i)S^{\pm}_{A_i},
\end{equation}
where
\begin{equation}
\gamma_i \equiv \prod^{k_i}_{j = 1}\lambda_{ij},\ \Phi_i \equiv \sum^{k_i}_{j = 1}\phi_{ij}.
\end{equation}
The most nontrivial $j$-th collision occurs when $\lambda_{ij} = 0$, and it results in $\gamma_i = 0$.  If $A_i$ does not collide with any environment qubit, then $\gamma_i = 1$ and $\Phi_i = 0$.  We remark that the above model is intrinsically discrete and has not been considered for analysis of robustness of $N$-partite entanglement.  However, Ziman {\em et al.} have shown that provided $0 < \lambda_{ij} < 1$ one can write down the master equation for a dephasing channel by performing some continuous-time approximation \cite{ZimanI, ZimanII}.  For nonzero $\lambda_{ij}$'s, the model thus could provide a microscopic description of the mechanism for a dephasing channel.  More significantly, we show that the model indeed yields new results quite different from previous investigations, especially when $\lambda_{ij} = 0$.  It is beyond the scope of this paper to discuss exactly how probable it is for $\lambda_{ij} = 0$ to occur.  However, only looking at Eq.(5) superficially, this may just happen often: for instance, when $\xi = \sigma^0/2$ and $X = \sin\theta\cos\phi\sigma^1 + \sin\theta\sin\phi\sigma^2 + \cos\theta\sigma^3$, where $\sigma^0$ is the $2 \times 2$ identity matrix, $\sigma^i$'s are the Pauli matrices, $0 \leq \theta \leq \pi$, and $0 \leq \phi \leq 2\pi$.

We start out by investigating the entanglement properties of $N$-qubit GHZ states, $|\Psi_{GHZ}\rangle_{A_1\cdots A_N} = (|0\cdots 0\rangle_{A_1\cdots A_N} + |1\cdots 1\rangle_{A_1\cdots A_N})/\sqrt{2}$ under this collisional decoherence.  $|\Psi_{GHZ}\rangle_{A_1\cdots A_N}$ has the density matrix
\begin{equation}
\rho^{GHZ}_{A_1\cdots A_N} = \frac{1}{2}
\left(\bigotimes^N_{i = 1}P^0_{A_i} + \bigotimes^N_{i = 1}S^+_{A_i} 
+ \bigotimes^N_{i = 1}S^-_{A_i} + \bigotimes^N_{i = 1}P^1_{A_i}\right).
\end{equation}
Applying the decoherence channel, Eq.(3), to every system qubit yields
\begin{equation}
\rho^{GHZ}_{A_1\cdots A_N} = \frac{1}{2}
\left(\bigotimes^N_{i = 1}P^0_{A_i} + \Gamma\bigotimes^N_{i = 1}S^+_{A_i} 
+ \Gamma^*\bigotimes^N_{i = 1}S^-_{A_i} + \bigotimes^N_{i = 1}P^1_{A_i}\right),
\end{equation}
with $\Gamma \equiv \left(\prod^N_{i = 1}\gamma_i\right)\exp\left(I\sum^N_{i = 1}\Phi_i\right)$.  Eq.(9) follows from Eq.(6).  Take partial transposition of the first $n$ qubits.  Since $P^{0, 1}_{A_i} \rightarrow P^{0, 1}_{A_i}$ and $S^{\pm}_{A_i} \rightarrow S^{\mp}_{A_i}$, we have
\begin{equation}
(\rho'^{GHZ}_{A_1\cdots A_N})^{T_{{\cal P}_1}} = 
\frac{1}{2}\left(\bigotimes^N_{i = 1}P^0_{A_i} + \Gamma\bigotimes^n_{i = 1} S^-_{A_i}\bigotimes^N_{i' = n + 1}S^+_{A_{i'}}
 + \Gamma^*\bigotimes^n_{i = 1} S^+_{A_i}\bigotimes^N_{i' = n + 1}S^-_{A_{i'}} + \bigotimes^N_{i = 1}P^1_{A_i}\right).
\end{equation}
Let $\chi \equiv (\rho'^{GHZ}_{A_1\cdots A_N})^{T_{{\cal P}_1}} - xI_{A_1\cdots A_N} = \left(\frac{1}{2} - x\right)\bigotimes^N_{i = 1}P^0_{A_i} - x\bigotimes^{N - 1}_{i = 1}P^0_{A_i} \otimes P^1_{A_N} - \cdots - x\bigotimes^{N - 1}_{i = 1}P^1_{A_i} \otimes P^0_{A_N} + \left(\frac{1}{2} - x\right)\bigotimes^N_{i = 1}P^1_{A_i} + \Gamma\bigotimes^n_{i = 1} S^-_{A_i}\bigotimes^N_{i' = n + 1}S^+_{A_{i'}} + \Gamma^*\bigotimes^n_{i = 1} S^+_{A_i}\bigotimes^N_{i' = n + 1}S^-_{A_{i'}}$.  $\det\chi = 0$ then yields $x^{2^N - 4}\left(\frac{1}{2} - x\right)^2\left(x^2 - \frac{1}{4}|\Gamma|^2\right) = 0$.  It is not difficult to deduce that, with respect to all bipartite cuts, $(\rho'^{GHZ}_{A_1\cdots A_N})^{T_{{\cal P}_1}}$ has the same negative eigenvalue given by
\begin{equation}
{\cal N}[\rho'^{GHZ}_{A_1\cdots A_N}] 
= -\frac{1}{2}\prod^N_{i = 1}\gamma_i 
= -\frac{1}{2}\prod^N_{i = 1}\prod^{k_i}_{j = 1}\lambda_{ij}.
\end{equation}
We note that a computable measure of entanglement associated with a bipartite cut ${\cal P}_1||{\cal P}_2$, the {\em negativity} \cite{Vidal} of $\rho'^{GHZ}_{A_1\cdots A_N}$ (or $\rho'^{GHZ}_{{\cal P}_1||{\cal P}_2}$) is given by $|{\cal N}[\rho'^{GHZ}_{A_1\cdots A_N}]|$. If $\lambda_{ij} = 1$ for all $i$ and $j$, then $|{\cal N}[\rho'^{GHZ}_{A_1\cdots A_N}]| = 1/2$.  As long as $0< \lambda_{ij} < 1$, $|{\cal N}[\rho'^{GHZ}_{A_1\cdots A_N}]|$ is nonzero, i.e., $\rho'^{GHZ}_{A_1\cdots A_N}$ remains $N$-partite entangled.  Obviously ${\cal N}[\rho'^{GHZ}_{A_1\cdots A_N}]$ depends on both $N$ and $k_i$.  When $k_i = K$ and $\lambda_{ij} = \lambda$ for all $i$ and $j$, we have
\begin{equation}
{\cal N}[\rho'^{GHZ}_{A_1\cdots A_N}] = -\frac{1}{2}\lambda^{KN}.
\end{equation}
Eq.(12) agrees with the results of analysis for the dephasing channel in \cite{Carvalho}.  For fixed $\lambda$ and $N$, $|{\cal N}[\rho'^{GHZ}_{A_1\cdots A_N}]|$ goes exponentially to zero as $K \rightarrow \infty$, and for fixed $K$ and $\lambda$, $\ln|{\cal N}[\rho'^{GHZ}_{A_1\cdots A_N}]|$ varies linearly with $N$.  However, we emphasize here that all it takes is merely one $\lambda_{ij} = 0$ to reduce ${\cal N}[\rho'^{GHZ}_{A_1\cdots A_N}]$ to $0$, and in this case, for $\rho'^{GHZ}_{A_1\cdots A_N}$ to become separable.  In this sense, we conclude that the $N$-qubit entanglement associated with a GHZ state is extremely fragile under collisional decoherence.

Next, we turn our attention to the case of $N$-qubit W states, 
$|\Psi_W\rangle_{A_1\cdots A_N} = 
(|0\cdots 01\rangle_{A_1\cdots A_N} + |0\cdots 010\rangle_{A_1\cdots A_N} + \cdots + |10\cdots 0\rangle_{A_1\cdots A_N})/\sqrt{N}$,
which has the density matrix
\begin{equation}
\rho^W_{A_1\cdots A_N} = \frac{1}{N}\left(\sum^N_{m = 1}\Pi_m
+ \sum_{1\leq m < m'\leq N}\Sigma^{+-}_{mm'} + \Sigma^{-+}_{mm'}\right),
\end{equation}
with $\Pi_m \equiv \bigotimes^{m - 1}_{i = 1}P^0_{A_i}\otimes P^1_{A_m}\otimes \bigotimes^N_{i = m + 1}P^0_{A_i}$ and $\Sigma^{\pm\mp}_{mm'} \equiv \bigotimes^{m - 1}_{i = 1}P^0_{A_i}\otimes S^{\pm}_{A_m}\otimes \bigotimes^{m' - 1}_{i = m + 1}P^0_{A_i}\otimes S^{\mp}_{A_{m'}}\otimes \bigotimes^N_{i = m' + 1}P^0_{A_i}$.  There are altogether $N(N - 1)$ terms in the second summation.  To any term with a pair $S^+_{A_m}$ and $S^-_{A_{m'}}$, there is a corresponding term with the pair $S^-_{A_m}$ and $S^+_{A_{m'}}$.  Applying the decoherence channel, Eq.(3), to every qubit yields
\begin{equation}
\rho'^W_{A_1\cdots A_N} = \frac{1}{N}\left[\sum^N_{m = 1}\Pi_m + \sum_{1\leq m < m'\leq N}
\left(\Upsilon_{mm'}\Sigma^{+-}_{mm'} + \Upsilon^*_{mm'}\Sigma^{-+}_{mm'}\right)\right],
\end{equation}
where $\Upsilon_{mm'} \equiv \gamma_m\gamma_{m'}\exp[I(\Phi_m - \Phi_{m'})]$.  As above, taking partial transposition of the first $n$ qubits, we have
\begin{eqnarray}
(\rho'^W_{A_1\cdots A_N})^{T_{{\cal P}_1}}
& = & \frac{1}{N}\left[\sum^N_{m = 1}\Pi_m
    + \sum_{1\leq m < m'\leq n}\left(\Upsilon_{mm'}\Sigma^{-+}_{mm'} + \Upsilon^*_{mm'}\Sigma^{+-}_{mm'}\right)\right. \nonumber \\
& & + \sum_{1\leq m \leq n < m'\leq N}\left(\Upsilon_{mm'}\Sigma^{--}_{mm'} + \Upsilon^*_{mm'}\Sigma^{++}_{mm'}\right) \nonumber \\
& & + \left.\sum_{n + 1\leq m < m'\leq N}\left(\Upsilon_{mm'}\Sigma^{+-}_{mm'} + \Upsilon^*_{mm'}\Sigma^{-+}_{mm'}\right)\right],
\end{eqnarray}
where $\Sigma^{\mp\mp}_{mm'} \equiv \bigotimes^{m - 1}_{i = 1}P^0_{A_i}\otimes S^{\mp}_{A_m}\otimes \bigotimes^{m' - 1}_{i = m + 1}P^0_{A_i}\otimes S^{\mp}_{A_{m'}}\otimes \bigotimes^N_{i = m' + 1}P^0_{A_i}$.  Similarly, we consider $\chi \equiv (\rho'^W_{A_1\cdots A_N})^{T_{{\cal P}_1}} - xI_{A_1\cdots A_N}$ and $\det\chi = 0$.  However, in contrast to Eq.(11), depending on the bipartite cuts, $(\rho'^W_{A_1\cdots A_N})^{T_{{\cal P}_1}}$ has negative eigenvalue given by
\begin{equation}
{\cal N}[\rho'^W_{A_1\cdots A_N}] = -\frac{1}{N}\sqrt{\left(\sum_{A_i \in {\cal P}_1}\gamma^2_i\right)
\left(\sum_{A_{i'} \in {\cal P}_2}\gamma^2_{i'}\right)}.
\end{equation}
And, the negativity associated with a bipartite cut ${\cal P}_1||{\cal P}_2$ is given by $|{\cal N}[\rho'^W_{A_1\cdots A_N}]|$.  If $\lambda_{ij} = 1$ for all $i$ and $j$, then $|{\cal N}[\rho'^W_{A_1\cdots A_N}]| = \sqrt{n(N - n)}/N$.  Eq.(16) agrees with Eq.(11) for $N = 2$.  As above, $\rho'^W_{A_1\cdots A_N}$ remains entangled so long as $0 < \lambda_{ij} < 1$.  However, ${\cal N}[\rho'^W_{A_1\cdots A_N}]$ clearly has a very different dependence on both $N$ and $k_i$ compared to ${\cal N}[\rho'^{GHZ}_{A_1\cdots A_N}]$.  When $k_i = K$ and $\lambda_{ij} = \lambda$ for all $i$ and $j$, we have
\begin{equation}
{\cal N}[\rho'^W_{A_1\cdots A_N}] = -\frac{1}{N}\sqrt{n(N - n)}\lambda^{2K},
\end{equation}
which reduces to $-\lambda^{2K}/2$ independent of $N$ if $n = N/2$.  On the other hand, when only $i = 1$ and $k_1 = K$, we have either
\begin{equation}
{\cal N}[\rho'^W_{A_1\cdots A_N}] = -\frac{1}{N}\sqrt{(n - 1 + \lambda^{2K})(N - n)},
\end{equation}
or
\begin{equation}
{\cal N}[\rho'^W_{A_1\cdots A_N}] = -\frac{1}{N}\sqrt{n(N - n - 1 + \lambda^{2K})},
\end{equation}
depending on if $A_1 \in {\cal P}_1$ or $\in {\cal P}_2$.  These remain nonzero even in the limit of infinite $K$.  Hence, the ``weakest links'' come from Eq.(17).  Again, Eq.(17) is consistent with the results of analysis for the dephasing channel in \cite{Carvalho}.   For fixed $\lambda$ and $N$, $|{\cal N}[\rho'^W_{A_1\cdots A_N}]|$ goes exponentially to zero as $K \rightarrow \infty$.  $|{\cal N}[\rho'^W_{A_1\cdots A_N}]|$ decreases as the number of qubits $N$ increases.  But, this decrease is solely due to the property of the original pure W state.  Therefore, modulo this effect, the entanglement decay rate is independent of $N$.  Here, we emphasize that in contrast to Eq.(11), Eq.(16) implies that when $\lambda_{ij} = 0$ or $\gamma_i = 0$ for one $A_i$, the state of the remaining $N - 1$ qubits $A_1, \cdots, A_{i - 1}, A_{i + 1}, \cdots, A_N$ is still $(N - 1)$-partite distillable entangled, even though $\rho'^W_{A_1\cdots A_N}$ is no longer $N$-partite distillable entangled.  Furthermore, it has to take $\gamma_i = 0$ for all $A_i$ to reduce the $N$-qubit state to a completely separable one.  We may thus conclude that the multipartite correlations of W states outperform those of GHZ states in terms of their robustness under collisional decoherence.

Lastly, we consider a linear cluster state of $N$ qubits \cite{Briegel}.  The state can be written in the form
\begin{equation}
|\Psi_C\rangle_{A_1\cdots A_N} = \sqrt{\frac{1}{2^N}}\bigotimes^N_{i = 1}(|0\rangle_{A_i}\sigma^3_{A_{i + 1}} + |1\rangle_{A_i})
\end{equation}
with the convention $\sigma^3_{A_{N + 1}} \equiv 1$.  For $N = 2$, applying the docherence channel, Eq.(3), to both qubits, we obtain the resulting state $\rho'^C_{A_1A_2}$ with negativity
\begin{equation}
{\cal E}[\rho'^C_{A_1A_2}] = \max\{\eta_{12}, 0\},
\end{equation}
where $\eta_{12} \equiv (\gamma_1\gamma_2 + \gamma_1 + \gamma_2 - 1)/4$.  When $\lambda_{ij} = 1$ for all $i$ and $j$, it gives $1/2$, which agrees with both Eqs.(11) and (16).  However, for $0 \leq \lambda_{ij} < 1$, it obviously yields different results.  This is despite the fact that $|\Psi_C\rangle_{A_1A_2}$ is up to a local unitary operation a maximally entangled Bell state \cite{Briegel}.  Consequently, supposing $\gamma_1 = \gamma_2 = \gamma$, the bipartite state $\rho'^C_{A_1A_2}$ becomes separable at $\gamma^{(2)}_{crit} = -1 + \sqrt{2}$.  In this sense, the bipartite linear cluster state is less robust compared to the corresponding GHZ or W states.  This fact that the robustness of a state can be affected by local operations has been discussed in \cite{Simon}.  Similarly, for $N = 3$ we have $\rho'^C_{A_1A_2A_3}$ with negativity associated with the bipartite cut $A_1 \in {\cal P}_1$ and $A_2, A_3 \in {\cal P}_2$ given by
\begin{equation}
{\cal E}[\rho'^C_{A_1(A_2A_3)}] = \max\{\eta_{12}, 0\}.
\end{equation}
By symmetry, we have with $\eta_{23} \equiv (\gamma_2\gamma_3 + \gamma_2 + \gamma_3 - 1)/4$,
\begin{equation}
{\cal E}[\rho'^C_{(A_1A_2)A_3}] = \max\{\eta_{23}, 0\}.
\end{equation}
But for the bipartite cut $A_1, A_3 \in {\cal P}_1$ and $A_2 \in {\cal P}_2$, we have instead
\begin{equation}
{\cal E}[\rho'^C_{(A_1)A_2(A_3)}] = \max\{\eta_{12}, \eta_{23}, \eta_{123}, 0\},
\end{equation}
where $\eta_{123} \equiv [(1 + \gamma_1)\gamma_2(1 + \gamma_3) - (1 - \gamma_1)(1 - \gamma_3)]/8$.  Eqs.(24) - (26) give $1/2$, which agrees with Eq.(11) when $\lambda_{ij} = 1$ for all $i$ and $j$.  This is again due to the fact that a tripartite linear cluster state is up to local unitary operation equivalent to a tripartite GHZ state \cite{Briegel}.  For $0 \leq \lambda_{ij} < 1$, and assuming $\gamma_1 = \gamma_2 = \gamma_3 = \gamma$, ${\cal E}[\rho'^C_{(A_1A_2)A_3}] = {\cal E}[\rho'^C_{A_1(A_2A_3)}] = 0$ at $\gamma^{(2)}_{crit}$ but ${\cal E}[\rho'^C_{(A_1)A_2(A_3)}] = 0$ at $\gamma^{(3)}_{crit} \approx 0.295598$.  We again conclude that the tripartite linear cluster state is less robust compared to the corresponding GHZ state.  The above cases exhibit two important properties generic to all $N$-partite linear cluster states.  First, among the negative eigenvalues of the partial transposed states across all possible bipartite cuts ${\cal P}_1$ and ${\cal P}_2$ are $\eta_{ij}$, $\eta_{ijk}$, etc.  Second, although $\gamma^{(3)}_{crit} < \gamma^{(2)}_{crit}$ it is nonzero, and the weakest links come from the $\eta_{ij}$'s.  Unlike the W states, $|\Psi_C\rangle_{A_1\cdots A_N}$ can thus fail to be $N$-partite distillable entangled for a range of nonzero values of $\gamma_i$.  Therefore, the linear cluster states are less robust compared to the W states, even though in this case it also has to take $\gamma_i = 0$ for all $A_i$ to reduce an $N$-qubit linear cluster state to a completely separable one.

In conclusion, we have shown that under collisional decoherence the multipartite entanglement associated with the GHZ, W and linear cluster states may not be robust.  In particular, an $N$-partite state can easily fail to remain $N$-partite distillable entangled.  Our results may seem contrary to some recent investigations.  However, it is not surprising, as this is the first time an inherently discrete model of decoherence has been considered for analysis of this kind.  It shows that the debate over if multipartite entanglement should be relatively common is far from over.  We end with the following remark: should there be some sort of mechanism that causes $\lambda_{ij} = |{\rm tr}[\xi X_{E_{ij}}]| = 0$ to occur very frequently, it could provide an answer to why multipartite entangled states are not prevalent in our macrscopic classical world.


\begin{thebibliography}{99}
\bibitem{CommentI} This agreement is truely remarkable given that these papers employ completely different measures of real multipartite entanglement.
\end{thebibliography}
\end{document}